\documentclass{speauth}

\usepackage{amssymb}
\usepackage{amsmath}
\usepackage{array}
\usepackage{amsfonts}\color[rgb]{0,0,0}
\usepackage{graphicx}
\usepackage{color}
\usepackage{version}
\usepackage{longtable}
\usepackage{supertabular}
\usepackage{comment}

\def\volumeyear{2017}

\def\figW{3.5in} 
\def\figsketch1{0.5\linewidth}
\def\figsketch2{0.75\linewidth}

\makeatletter
\def\hlinewd#1{%
\noalign{\ifnum0=`}\fi\hrule \@height #1 %
\futurelet\reserved@a\@xhline}
\makeatother

\renewcommand{\journalname}{}
\renewcommand{\journalabb}{}

\begin{document}

\runningheads{L.~Le~Guen et al.}{Heat convection and radiation in flighted rotary kilns: a minimal model}

\title{{HEAT CONVECTION AND RADIATION IN FLIGHTED ROTARY KILNS: A MINIMAL MODEL}}
 \author{Laur\'edan Le Guen, 
 Maxime Piton,
 Quentin H\'enaut,
 Florian Huchet,
 and Patrick Richard\corrauth}
\address{LUNAM Universit\'e, GPEM, IFSTTAR, site de Nantes,  Route de Bouaye, CS4 44344 Bouguenais Cedex, France}
\begin{abstract}
We propose a minimal model aiming to describe heat transfer  between particles ({i.e.} grains) and gases in a model of flighted rotary kilns.
It considers  a channel in which a convective gas interacts with a granular suspension and a granular bed.  Despite its simplicity it captures the main experimental findings in the case of  dilute suspension of heavy grains 
typical of what can be observed in many industrial rotary kilns.
Energy balance between each phase takes  into account the main heat transfer mechanisms between the transverse granular motion and the convective gas. 
In the absence of  radiation heat transfer,  the model predicts exponential variations of the temperatures characterized by a length which depends on the granular and gas heat flow rates  as well as on the exchange areas.  
When radiation is taken into account, the model can be solved numerically. 
For this case, the temperature variations can be fitted by stretched exponentials whose parameters are found to be independent of the studied phases. Finally, an efficiency criterion is proposed to optimize the length of the system.\\

This is a reprint of an article accepted for publication in The Canadian Journal of Chemical Engineering, \copyright copyright (2017)


\end{abstract}


\maketitle

\vspace{-6pt}

\section*{INTRODUCTION}
In recent years, there has been a great deal of interest in the understanding of the behaviour of granular materials as they are used in a large number of engineering processes and present in many geophysical systems. 
Accordingly, many progresses have been made on the theoretical description of these systems and on the understanding of the physical mechanisms that govern their behaviour when they are exhibiting a gaseous, a liquid or a solid-like behavior. For instance, dilute granular flows are well described by a kinetic theory~\cite{jenkins1997,Delannay2007} and unidirectional dense flows properties are well captured by a Druker-Parger-like rheology~\cite{Jop_Nature_2006} or by extensions of the aforementioned kinetic theory~\cite{Jenkins_granularmatter_2012}.
The behavior of such materials in the vicinity of jamming~\cite{Papanikolaou_PRL_2013,Daniels_PRE_2012,Bi_Nature_2011,PicaCiamarra2011,PicaCiamarra_SM_2012,McNamara2009,Crassous_JSTAT_2008}, close to destabilization~\cite{Nerone2003,Zaitsev2008} or submitted to aging processes~\cite{Ribiere2005b,KiesgendeRichter2010,Espindola_PRL_2012,Zaitsev_PRL_2014} is also the subject of active current research.\\
{Heat transfer in such systems~\cite{Vargas_PRE_2007,huetter_JGR_2008} and in particularly in gas-grains mixtures~\cite{Shi_ChemEngScience_2008,Chen_IntJMassTrans_2008,Zhou_AIChE_2009,Boateng_IntJournHeatMassTrans_1996,Caputo_ApplThermEng_2011} is an important physical phenomenon which may govern  a wide variety of natural systems (e.g. volcanic eruption)  and industries (rotary kilns devoted to asphalt or cement production~\cite{Caputo_ApplThermEng_2011}, torrefaction of biomass~\cite{Colin_ChemEngResDes_2015},  fertilizers production, waste treatment, manufacturing of nuclear fuel~\cite{Debacq_Powdertechnology_2013}\ldots).}  
{This phenomenon is } quite complex for several reasons~\cite{Mastorakos_ApplMathModel_1999}. Among them, 
we can mention that 
heat transfer may involve three different physical phenomena: (i) conduction (ii) convection and (iii) radiation.
The case of conduction is also quite complicated {by itself} since it is influenced by the nature of the contacts between grains which 
is controlled by the nano-asperities present at the surface of the grains~\cite{Zaitsev_PRL_2014}.\\  
Recent and important {progresses have} been made in the numerical simulation of fluid interacting with moving particles using Discrete Element Methods (DEM) coupled to computational fluid dynamics (CFD) --see for example \cite{Zhou_AIChE_2009,Lomine2013,Topin_PRL_2012,Mutabaruka_PRE_2014} or ~\cite{Shi_ChemEngScience_2008,Hobbs_IntJCompFD_2009,Wen_JEngMech_2014} for systems similar to ours--.
However,  their practical use 
{to simulate industrial facilities} is partially limited  {because they require important computational resources.}
Therefore, it is still necessary to develop physical models~\cite{Sheehan_ChemEngSc_2005,Hamawand_TCJChEng_2014,Tsamatsoulis_TCJChEng_2014,Ajayi_IndEngChemRes_2015} capable to reproduce and predict the heat transfer between a gas and a granular medium.\\
{Here, we (i) present  a minimal physical model   which aims to quantify the heat transfer 
{in the case of wall-bounded gas flow through dilute suspension of heavy grains} and (ii) use this model to analyze the thermal efficiency of rotary kilns}. 
In particular, we discuss the relative importance of radiation heat transfer with respect to 
{convection}  and show that, the presented model captures the main experimental findings 
{obtained in industrial facilities}.\\

\section*{ROTARY KILN}\label{sec:kiln}
\begin{figure}[htbp]
\begin{center}
\includegraphics*[width=\figW]{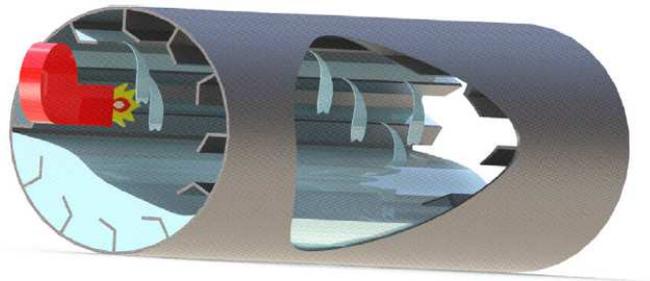}
\caption{Sketch of a typical rotary kiln. The heat is provided by a direct fire natural burner. Shape metal slats called blades or flights are attached to the interior surface of the kiln. They lift particles located at the bottom of the kiln to the gas. The kiln is inclined (a few degrees) with respect to the horizontal.}\label{fig:drum}
\end{center}
\end{figure}

Granular heat transfer modeling has a great importance in the material processing field. For example, rotary kilns are  the most popular equipment dedicated to the drying, heating and coating of granular materials. A better understanding of the physical mechanisms~\cite{Schlunder_CEPI_1984,Herz_IJHMT_2012,Elattar_TCJChEng_2016} occurring within  a rotary kiln is therefore crucial to optimize the industrial plants~\cite{LeGuen_ExTFS_2014} in the context of sustainability, efficient energy use and reduction of toxic gases like polycyclic aromatic hydrocarbons.\\
Here, we aim to model such a typical industrial equipment. A rotary kiln is a cylinder 
slightly tilted (a few degrees) with respect to the horizontal and which rotates around its axis (Figure~\ref{fig:drum}).\\
The material to be processed (typically grains) is fed into the upper end of the cylinder (the inlet). Due to gravity and the rotation of the kiln, the material gradually moves down towards the lower end (the outlet). In the case of grains, they are heated by a  hot gas passing along the axis of the cylinder from the upper end to the lower end (co-current kiln) or from the lower end to the upper end (counter-current kiln). 
The hot gases are generally provided by an external furnace, or by a flame inside the kiln.\\
In such a system, the grains are initially located at the bottom of the cylinder. 
However, L-shaped baffles {(also called flights)} located along the cylinder parallel to its axis on its inner diameter lift grains from the bed, bring them up to the top of the cylinder where they fall down due to gravity~\cite{Sheehan_Powdertech_2010}. 
{A granular curtain is thus created~\cite{Sheehan_Powdertech_2010}. It }
ensures a proper mixing
between grains and hot gases, and so enhances the heat exchange. 
It is thus convenient to consider that the physical system is made of four phases: (i) the gas, (ii) the dense granular bed, (iii) a dilute granular suspension {i.e.} a granular curtain and (iv) the walls bounding the system, and to study the heat exchange between those four phases.\\ 
Although the  analysis presented in this work can be easily extended to the counter-current case, we will restrict ourselves to the co-current configuration.
This choice may be surprising because it is reasonable to think that the maximum amount of heat transfer that can be obtained is more important with a counter-current kiln than with its co-current counterpart.
Indeed, the former maintains a slowly declining temperature gradient whereas in the latter, the strong thermal difference at the inlet induces a higher initial gradient which falls off quickly, leading to {potential heat loss.} 
However, co-current rotary drum are still predominantly used in the industry for practical reasons: they are more easy to build, the fuel cost is nowadays not expensive enough to justify the replace old co-current kilns by new counter-current ones.

\section*{HEAT TRANSFER MODEL}\label{sec:model}

\begin{figure}[htbp]
\begin{center}
\includegraphics*[width=\figW]{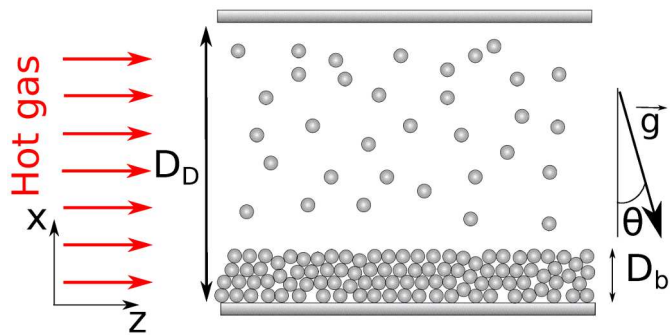}
\caption{Sketch of the system. Two horizontal walls are separated by a distance $D_D$. 
A hot gas flows in the system in parallel with the walls. A granular bed (height $D_b$) rests at the lower wall. L-shaped baffles (not shown) lift grains from the granular bed to the upper wall leading to an homogeneous suspension of grains which falls down due to gravity. The whole is inclined at an angle $\theta$ with respect to the horizontal, leading to a global motion of the grains along the drum's axis.}\label{fig:sketch_bis}
\end{center}
\end{figure}

In our study case, the rotary kiln is approximated by a drum where heat is exchanged between different phases that will be defined below. The drum is a cylinder made in stainless steel with an inner diameter, $D_D$ a length, $L_D$ and a thickness, $d_D$. 
The convective hot gas flows from one boundary of the drum (the inlet) to the other (the outlet) along its axis. To simplify the problem we model the drum by 
two infinite  and insulated horizontal walls separated by a distance $D_D$ (see Figure~\ref{fig:sketch_bis}). \\
{This simplification neglects the action of rotation which is justified by the very small rotation speed used in industrial facilities.}
A gas is convected in parallel with those two walls and between the inlet of the system, defined by $z=0$, and the outlet, defined by $z=L_D$.
A packing of grains (diameter $d_s$, volumic mass density $\rho_s$, thermal conductivity $k_s$) whose height is denoted by $D_b$, is located on the lower wall. The whole is inclined with respect to the horizontal by a small angle $\theta$ (few degrees). 
 We set the vertical origin at the level at the lower wall ($x=0$). The upper wall is therefore located at $x=D_D$.
The action of the L-shaped baffles is implicitly taken into account by the presence of a uniform suspension  ({i.e.} the volume fraction of the suspension depends neither on $x$ nor on $z$) of the same type of grains which interact with the convected gas.  
In the granular suspension, the net transfer of thermal energy is potentially enhanced by an effective conduction from collisions between grains.\\
{However, in most practical cases the collision times between grains are too small to lead to a significant heat transfer between particles~\cite{Sun_IntHeatMassTransfer_1988} and the volume fraction of the granular suspension is small enough to consider that the average number of collisions per grain is very weak.
Moreover, we also assume that the grain size is large enough to neglect the velocity fluctuations of grains.} In other words, the granular temperature and thus the self-diffusion of grains are negligible. Our model is therefore fundamentally different from that of Chen and Louge~\cite{Chen_IntJMassTrans_2008} and clearly not valid for highly agitated granular systems where the granular temperature is far from being negligible.\\
{It should be pointed out that such assumptions are justified by practical reasons. Our aim is not to derive a general and complex model that captures the behavior of a large variety of configurations but to focus on an applied point of view by deriving a minimal model devoted to thermal rotary kilns which are currently and widely used in the industry. Of course, we will also study the predictions and the limits of the aforementioned model.} 
Another assumption consists in  neglecting the momentum transfer between the gas and the grains. In other words the grains are too large and heavy to be influenced by the gas (Stokes number $\gg 1$), 
{and the granular suspension is dilute enough ($<10 \%$) to neglect its influence on the gas.}
The only interaction between grains and gas is therefore related to the heat transfer. The motion of grains is induced by gravity through the inclination of the kiln and through the action of baffles, whose size is negligible  compared to that of the drum. 
{The former and the latter assumptions prevent the use of our work to model fluidized beds~\cite{Zhou_AIChE_2009,Chen_powdertechnology_2005} of other systems where the grains are highly agitated and where fluid grains interactions govern significantly the dynamics.}

%

We also assume that the Biot number of grains $B_i = h(d_s/2)/k_s$, where $h$ is the heat transfer coefficient, is small with respect to $0.1$. Thus, their internal temperature can be considered as uniform. 
{Finally our  last assumption is the following: for a given $z$-position along the axis of the drum, the grains belonging to the granular bed and to the granular suspension have the same temperature, $T_s$. Therefore the heat transfer between grains of the suspension and those belonging to the bed is not considered.} Such an assumption, which has been widely used previously in the literature~\cite{Boateng_IntJournHeatMassTrans_1996,Patisson_MetMatTransB_2000} is an important simplification. Yet, in the case of a rotary kiln, the presence of baffles ensures the solid phase motion in order to renew the exchanges surfaces~\cite{Chaudhuri_powdertechonolgy_2010} and thus a reasonable homogeneity of the grain temperature. The relevancy of this simplification has been recently confirmed by numerical simulations coupling DEM with CFD~\cite{Hobbs_2015} which reports, in a counter-current kiln, a reasonably weak variation of temperature of the solid phase (the maximal standard deviation is approximately $20$ K for an average temperature of approximately $320$ K). Moreover such an assumption is fully consistent with our aforementioned aim, that is, to derive the minimal model which can be used at the industrial scale. Our model is thus therefore a reasonable alternative to heavier approaches like numerical simulations coupling DEM and CFD. Thus, the grain temperature $T_s$ is assumed to be only a function of $z$. Similarly, the gas and the wall temperatures (respectively $T_g$ and $T_w$) are also assumed to depend only on $z$. 

In such a system, several types of 
{convective} heat transfer should be taken into account:
\begin{itemize}
\item The convective heat transfer between the granular bed and the gas given by: $h_{bg}\,dS_{bg}\left(T_{g}-T_s\right)$, where $h_{bg}$ is the convective exchange coefficient between the granular bed and the gas, and 
$dS_{bg}$, the differential surface area between the granular bed (or more precisely its free surface) and the gas. 

\item The convective transfer between the gas and the granular suspension ({i.e.} the granular curtain): $h_{cg} dS_{cg}\left(T_g-T_s\right)$,
 where $h_{cg}$ and $dS_{cg}$ are respectively the convective exchange coefficient and 
 the differential surface area between the granular suspension and the gas.

\item The convective transfer between the granular bed and the contacting wall: $h_{sw}\, dS_{sw}\left(T_w-T_s\right)$.
Similarly, $h_{sw}$ is the convective exchange coefficient and 
$dS_{sw}$, the differential surface area between the granular bed and the inner contacting walls.
\item The convective transfer between the gas and the walls: 
$h_{gw}\,dS_{gw}\left(T_w -T_g \right),$
where $h_{gw}$ is the gas-wall convective exchange coefficient and $dS_{gw}$ the corresponding differential surface area.
\end{itemize}
{From the differential surfaces $dS_{ij}$ we can define length $l_{ij}$ by $l_{ij}=dS_{ij}/dz$ where $dz$ is 
the differential of $z$.
}
{The heat conduction between grains belonging to the granular bed is assumed to be negligible.}
{It should be pointed out that, by assuming a constant grain diameter whereas real systems are polysized, we underestimate the heat exchange between the convected gas and the granular suspension. Yet, in view of industrial applications 
such an underestimation is better than an overestimation.}
The exchange surfaces between each phase are determined following the work of Piton et al.~\cite{Piton_AppliedThermalEng_2015}. 
{Using the same kiln geometry than ours and results from the literature, they provide expressions for the heat transfer coefficients which are empirical or semi-empirical. For the sake of clarity we also reported those expressions in   tables~\ref{tab:hbg}, \ref{tab:hcg}, \ref{tab:hgw} and \ref{tab:hsw}.}
As we will show below, one of the key parameter is the 
fraction of grains in the granular suspension, $\alpha_{cg}$, {i.e.} the volume of the grains belonging to the suspension divided by the total volume of grains. 
The latter quantity is directly linked to the number of grains contained in the granular suspension per unit of length $dN/dz$ through: 
\begin{equation}
\alpha_{cg}=\frac{2}{3}\frac{dN}{dz}\frac{d_s^3}{D_D^2\, F_t},
\end{equation}
where $F_t$ is the ratio of the volume of grains to the volume of the drum, {i.e.} the kiln volume fraction.
Then, the convective heat transfer coefficients are calculated from correlations available in the literature~\cite{Piton_AppliedThermalEng_2015}.\\ 
Thermal radiation is modeled using the following heat transfers:
\begin{itemize}
\item between the gas and the granular suspension: $dS_{cg} E_{cg} \left(T_g^4-T_s^4 \right)$,
\item between the gas and the free surface of the granular bed: $dS_{bg}E_{bg} \left(T_g^4-T_s^4\right)$,
\item between the gas and the wall of kiln: $dS_{gw}E_{gw} \left(T_w^4-T_g^4\right)$,
\item between the granular bed and the contacting wall of the kiln $dS_{sw}E_{sw} \left(T_w^4-T_s^4\right)$.
\end{itemize}
In the latter expressions, the  quantities $E_{ij}$ with $i,j=\{c,b,g,w\}$ are the radiative heat transfer coefficients defined as 
$E_{i,j}=\sigma \varepsilon_{ij}$ where $\sigma$ is the Stefan-Boltzmann constant and $\varepsilon_{ij}$ the corresponding emissivity.
The heat exchange surfaces $dS_{ij}$ are those used in the convective heat transfer. 

So, at equilibrium, 
{using the aforementioned types of heat transfer, the energy balances (i.e. the first law of thermodynamics) for the solid 
and gas 
phases as well as the insulated wall condition 
are respectively given by: } 
\begin{eqnarray}
\delta \dot Q_s &= &\dot{m_s} \,C_{p,s}\,dT_s \nonumber\\
 & = &\left(h_{bg}\,dS_{bg}  \nonumber
 + h_{cg}\,dS_{cg}\right) \left(T_g -T_s\right) \nonumber \\
 & & + E_{gs} \left(dS_{bg} + dS_{cg}\right) \left(T_g^4 -T_s^4\right) \nonumber\\ 
 & &  + h_{sw}\,dS_{sw}\left(T_{w}-T_s\right) \nonumber\\
& & + E_{sw} \left(dS_{bg} + dS_{cg}\right)\left(T_w^4 -T_s^4\right),\label{eqn:Ts_rad}\label{eqn:Eq1}
\end{eqnarray}
\begin{eqnarray}
\delta \dot Q_g &= &\dot{m_g}\,C_{p,g}\,dT_g \nonumber\\
 & = &\left(h_{bg}\,dS_{bg} + h_{cg}\,dS_{cg}\right)\left(T_s -T_g\right)\nonumber \\ 
& & +E_{gs} \left(dS_{bg} + dS_{cg}\right) \left(T_s^4 -T_g^4\right) \label{eqn:Eq2} \nonumber\\
& & +h_{gw}\,dS_{gw}\left(T_{w}-T_g\right)\nonumber\\
& & +E_{gw} dS_{gw}\left(T_w^4 -T_g^4\right),\label{eqn:Tg_rad}
\end{eqnarray}
and
\begin{eqnarray}
h_{sw}\,dS_{sw}\left(T_s -T_w\right) + 
E_{sw} \left(dS_{bg} +   
dS_{cg}\right)\left(T_s^4 -T_w^4\right) \nonumber\\ \,+h_{gw}\,dS_{gw}\left(T_g -T_w\right) + E_{gw} dS_{gw}\left(T_g^4 -T_w^4\right) = 0. \label{eqn:Eq3}
\end{eqnarray}
{In the latter equations,  $\delta \dot Q_i$ (with $i=s,g$ respectively for the solid and the gas phase) is the heat transfer per unit time (the dot notation is used for the time derivative), $\dot{m_i}$  the mass flow rate of the phase $i$, and  $C_{p,i}$  the corresponding heat capacity.\\}
The set of Equations~(\ref{eqn:Eq1}), (\ref{eqn:Eq2}) and (\ref{eqn:Eq3})  composes a system with the following boundary values: $T_s\left(0 \right)=T_{s,0}$ and $T_g\left(0 \right)=T_{g,0}$ are respectively  the inlet temperature of solids ({i.e.} grains) and the inlet temperature of hot gases. 

\section*{ANALYTICAL RESOLUTION IN THE CASE OF NEGLIGIBLE THERMAL RADIATION}\label{sec:analytical}
The analytical resolution of previous model is not possible due to the presence of $T^4$-terms. 
However, neglecting the heat transfer by radiation leads to a classical system of equations which can be solved analytically. 
The first law of thermodynamics applied to the grains and to the gas  gives the respective heat flow rates :
\begin{eqnarray}
\dot{m_s} \,C_{p,s}\,dT_s&=&h_{bg}\,dS_{bg}\left(T_g -T_s\right)\nonumber\\
 & & +h_{cg}\,dS_{cg}\left(T_g-T_s\right) \nonumber\\
 & &\,+h_{sw}\,dS_{sw}\left(T_{w}-T_s\right),\label{eqn:eqdiffTs}
\end{eqnarray}
and
\begin{eqnarray}
\dot{m_g}\,C_{p,g}\,dT_g&=&h_{bg}\,dS_{bg}\left(T_s -T_g\right) \nonumber\\
& &+h_{cg}\,dS_{cg}\left(T_s-T_g\right) \nonumber\\
 & &+h_{gw}\,dS_{gw}\left(T_{w}-T_g\right).\label{eqn:eqdiffTg}
\end{eqnarray}
Since the system is insulated, the wall is in thermal equilibrium with the gas and the solid phase in such way that the energy balance  is given by :
$$h_{sw}\,dS_{sw}\left(T_{w}-T_s\right)+h_{gw}\,dS_{gw}\left(T_{w}-T_g\right)=0.$$
The latter equation leads to the identification of the inner wall temperature:
\begin{equation}
T_{w}=\frac{h_{sw}\,dS_{sw}T_s+h_{gw}\,dS_{gw}\,T_g}{ h_{sw}\,dS_{sw}+h_{gw}\,dS_{gw}}.\label{eqn:eqTw}
\end{equation}
Combining Equations~(\ref{eqn:eqdiffTs}), (\ref{eqn:eqTw})  and (\ref{eqn:eqdiffTg}) and using $dS_{i,j}=dz\,l_{i,j}$ and 
$\Psi_{ij}=h_{ij}\,l_{ij}$, with $i,j=\{c,g,w,b\}$ we obtain the following first order differential system 

$$
\begin{array}{ccc}
\displaystyle{
\frac{d\,T_g}{dz}}&=&\displaystyle{-\left[ \frac{\Psi_{bg}+\Psi_{cg}+\Psi_{gw}}{\dot m_g C_{p,g}} -
 \frac{\Psi_{gw}^2}{(\Psi_{sw} + \Psi_{gw})\dot m_g C_{p,g}}\right]T_g}\\
 &\ & \displaystyle{+\left[\frac{\Psi_{gw} + \Psi_{cg}}{\dot{m_g}C_{p,g} }+\frac{\Psi_{gw}\,\Psi_{sw}}{\left(\Psi_{sw} + \Psi_{gw} \right)\dot{m_{g}}C_{p,g}}\right]T_s,}\\
\displaystyle{\frac{d\,T_s}{dz}}&=&\displaystyle{-I\left[ \frac{\Psi_{gw} + \Psi_{cg} +\Psi_{gw}}{\dot m_g C_{p,g}} - \frac{\Psi_{gw}^2}{(\Psi_{sw}+\Psi_{gw})\dot m_g C_{p,g}}\right]T_s}\\
& \ &\displaystyle{ + I\left[\frac{\Psi_{bg}+\Psi_{cg}}{\dot{m_g}C_{p,g} }+\frac{\Psi_{gw}\,\Psi_{sw}}{\left(\Psi_{sw}+\Psi_{gw} \right)\dot{m_{g}}C_{p,g}}\right]T_g.}
\end{array}
$$
where $I$ is a dimensionless number defined as ${\dot{m_g}C_{p,g}}/{\dot{m_s}C_{p,s}}$.
After resolution, the latter system leads to:
\begin{equation}{\left\{
\begin{array}{ccc}
T_g &=& \displaystyle{\frac{T_{g,0}-T_{s,0} }{I+1 }\exp\left(-z/\Lambda \right)+\frac{T_{g,0}I+T_{s,0}}{I+1},}\\
T_s&=& \displaystyle{-I\frac{T_{g,0}-T_{s,0} }{I+1 }\exp\left(-z/\Lambda\right)+\frac{T_{g,0}I+T_{s,0}}{I+1}.}
\end{array}
\right.}\label{eqn:sol_exp}\end{equation}
In those solutions,   $\Lambda$ is a characteristic length defined as:
\begin{equation}
\Lambda =\frac{E}{I+1}\frac{1}{A+B+CD/(D+C)}\label{eqn:Lambda}
\end{equation}
with $A=h_{cg}l_{cg}$, $B=h_{bg}l_{bg}$, $C=h_{sw}l_{sw}$, $D=h_{gw}l_{{gw}}$ and  $E=\dot{m_g}C_{p,g},$
Thus, in the absence of thermal radiation 
{the temperature profiles along the drum} are given by exponential functions.
{In the remainder of the paper, we will consider the following nominal case. Grains are spheres of diameter $d_s=0.005\,\rm{m}$ and the curtain density is characterized by  $\alpha_{cg}=3\%$. The solid mass flow rate is $\dot m_s = 33.98\,\rm{kg.s}^{-1}$.} The gas  is defined by its mass flow rate $\dot m_g=3.74\,\rm{ kg.s}^{-1}$ (unless otherwise specified), its  density $\rho_g = 0.84\,\rm{kg.m}^{-3}$ and its viscosity $\mu_g = 3.59\,10^{-5}\,\rm{Pa.s}^{-1}$.  
{For the sake of simplicity, we do not take into account the variations of the two latter quantities with the temperature. However, in our model, they only influence $h_{cg}$ and $h_{gw}$ through the Reynolds number. Yet, in presence of radiation, the variations of $\mu_g$ and $\rho_g$ are really important only when radiation is negligible, {i.e.} for a gas temperature below $600\rm{ K}$. Within this range temperature, the effect on the heat transfer coefficients is negligible.}
The  convective heat transfer coefficients corresponding to the nominal case are given by the following values (see~\cite{Piton_AppliedThermalEng_2015}):
$h_{bg}=102.83~\rm{W.m}^{-2}\rm{.K}^{-1}$,
$h_{cg}=112.80~\rm{W.m}^{-2}\rm{.K}^{-1}$,
$h_{gw}=35.23~\rm{W.m}^{-2}\rm{.K}^{-1}$ and 
$h_{sw}=242.96~\rm{W.m}^{-2}\rm{.K}^{-1}$.
Note that, those values are obtained by semi-empirical relations (see tables~~\ref{tab:hbg}, \ref{tab:hcg}, \ref{tab:hgw} and \ref{tab:hsw}) obtained from the study of 
industrial rotary kilns used to process various types of granular matter~\cite{Piton_AppliedThermalEng_2015}
and that the heat transfer coefficients depend on both $\alpha_{cg}$ and $\dot m_g$. 
Although we focus here on the case of asphalt plants, our approach is valid for any type of granular material as long as the assumption of the model are valid. It should be the case, for example, 
for kilns used to dry fertilizers. 

For the nominal case, the exchange lengths, which  depend on the grain distribution in the kiln cross-section (see table~\ref{tab:l} and \cite{Piton_AppliedThermalEng_2015}), are given by 
$l_{bg}=2.320~\rm{m}$,
$l_{cg}=9.71~\rm{m}$,
$l_{gw}=3.55~\rm{m}$ and 
$l_{sw}=1.79~\rm{m}$.
Those values depend on $\alpha_{cg}$ and will also be modified accordingly when the effect of that parameter will be studied.
\begin{table}[htbp]
\begin{center}
\caption{The coefficient $h_{bg}$ depends on  $A_g$ (the cross section of ''hot gas'' phase) and  $\dot m_g$ (the mass flow
rate of hot gases in the kiln). The constants  $H_{bg}$, $\sigma_{bg}$ and $\eta_{bg}$ are respectively equal to 
$0.4~\mbox{W.}\mbox{m}^{-2}\mbox{.K}^{-1}$, $3600~\mbox{kg}^{-1}\mbox{.m}^2\mbox{.s}$ and $0.62$. 
}\label{tab:hbg}
\begin{tabular}{|>{\centering\arraybackslash}m{.12\textwidth} | >{\centering\arraybackslash}m{.68\textwidth} |}
\hline
\multicolumn{2}{|c|}{Convective transfer between the gas and the granular bed}\\
\hline
\vspace*{0.4cm}Coefficient\vspace*{0.4cm}&$h_{bg}=H_{bg} \displaystyle{\left({\sigma_{bg}\dot m_g  }/{ A_g}\right)^{\eta_{bg}} }$ \\
\hline
Validity range& $h_{bg} \in [50,100]~\mbox{W.m}^{-2}\mbox{K}^{-1}$~\cite{Gorog_MetallTransB_1982}\\
\hline
Sketch&\vspace{0.1cm} \includegraphics*[width=0.5\linewidth]{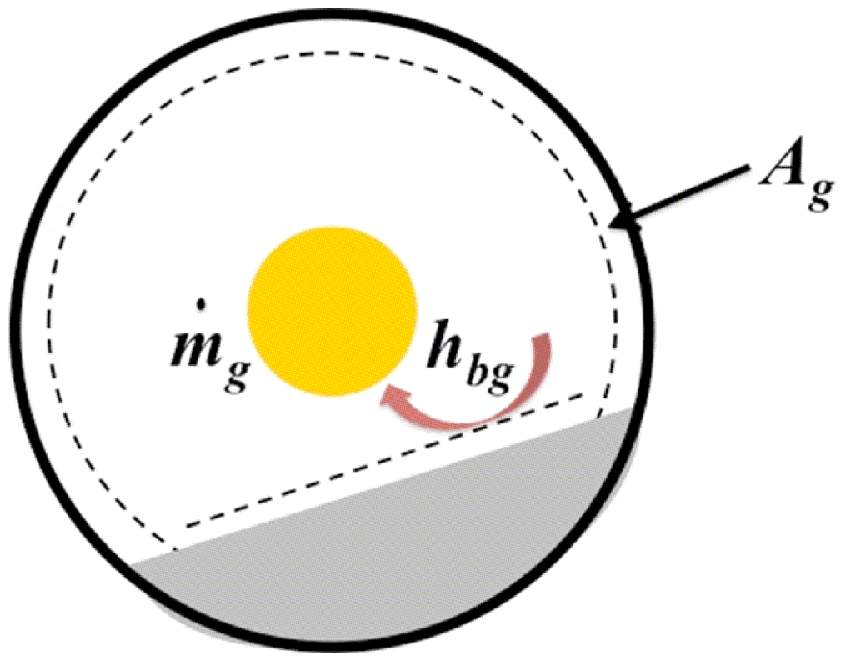}\\ 
\hline
\end{tabular}
\end{center}
\end{table}

\begin{table}[htbp]
\begin{center}
\caption{The coefficient $h_{cg}$ depends on 
$k_s$, the
thermal conductivity of solids, $d_s$ the grain size, $P_r$ the Prandtl number {i.e.} the ratio of 
kinematic viscosity to the thermal diffusivity and the particle Reynolds number $R_{ep}$ defined as $v_g d_s /\nu_g$, where $v_g$ and $\nu_g$ are respectively the gas velocity and the gas kinematic viscosity.
The coefficients $A_{cg}$, $B_{cg}$, $a$, $b$ and $c$ depend on the range of the particle Reynolds number $R_{ep}$~\cite{Li_PowderTech_2000}.
}\label{tab:hcg}
\begin{tabular}{|>{\centering\arraybackslash}m{.12\textwidth} | >{\centering\arraybackslash}m{.226\textwidth} |>{\centering\arraybackslash}m{.226\textwidth} |>{\centering\arraybackslash}m{.226\textwidth} | }
\hline
\multicolumn{4}{|c|}{Convective transfer between the gas and the granular suspension}\\
\hline
\vspace*{0.4cm}Coefficient\vspace*{0.4cm}&\multicolumn{3}{c|}{$h_{cg}=\displaystyle{\frac{k_s}{d_s}\left[2+\left(A_{cg}\,R_{e,p}^a+B_{cg}\,R_{e,p}^b\,P_r^c\right)\right]}$} \\
\hline
Validity range&  if $R_{e,p} < 200$, $A_{cg}=0$, $a=0$, $B_{cg}=0.6$, $b= 1/3$ and $c=1/3$ &
if $200 < R_{e,p} < 1500$, $A_{cg}=0.5$, $a=0$, $B_{cg}=0.02$, $b=0.8$ and $c=1/3$ &
if $R_{e,p}> 1500$, $A_{cg} = 4.5\times 10^{-5}$; $a=1.8$; $B_{cg}=0$, $c=0$, $d=0$
\\
\hline
Sketch&\vspace{0.1cm} \includegraphics*[width=0.75\linewidth]{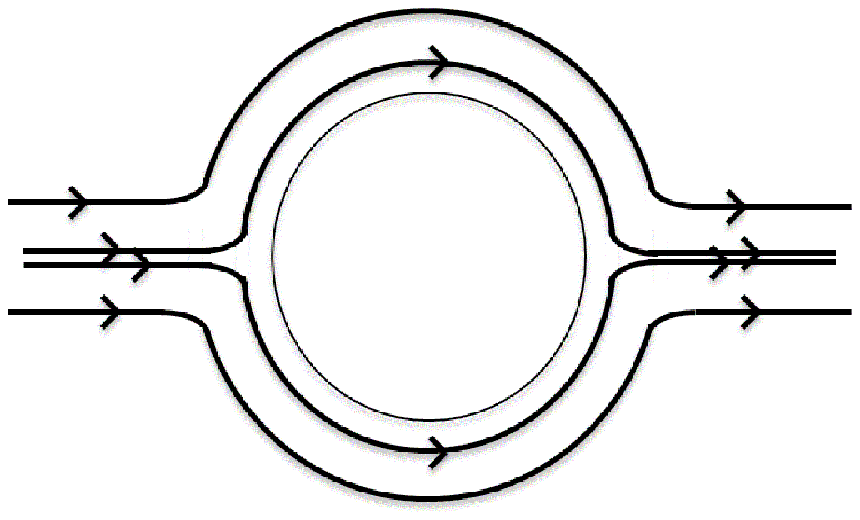}& 
\vspace{0.1cm} \includegraphics*[width=0.75\linewidth]{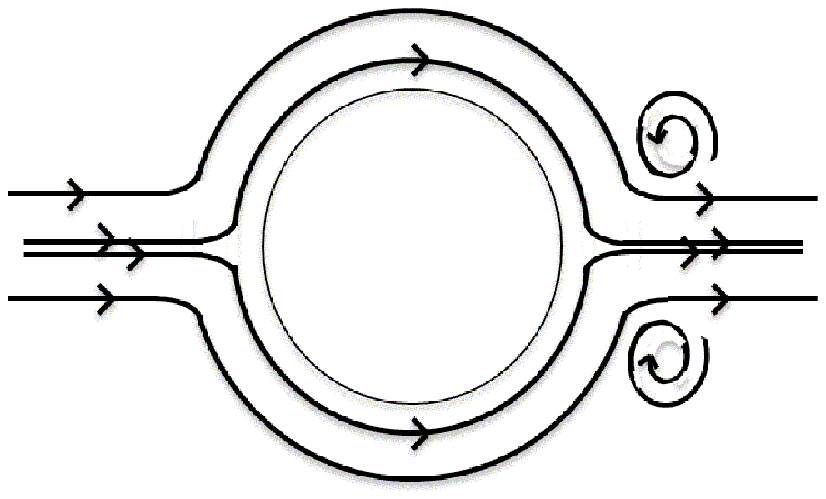}&
\vspace{0.1cm} \includegraphics*[width=0.75\linewidth]{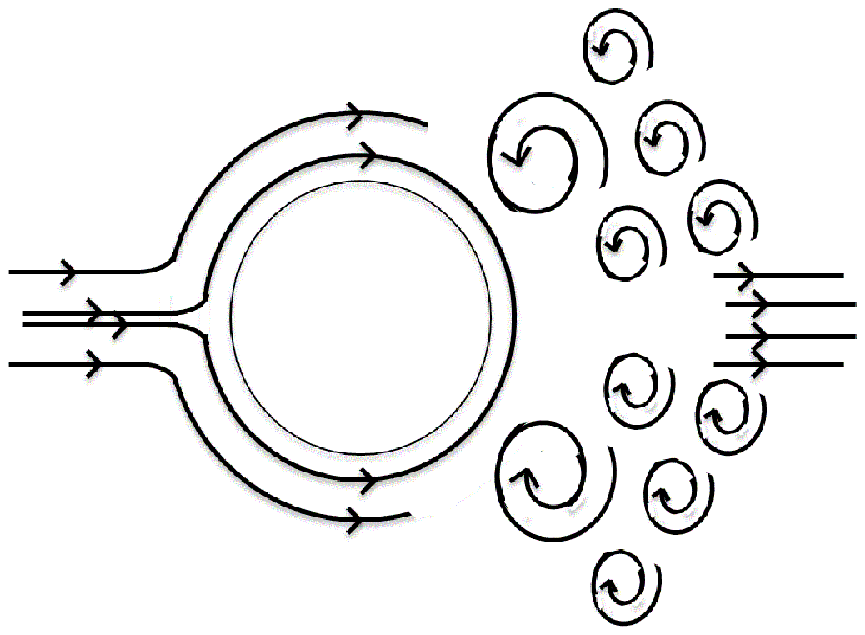}\\
\hline
\end{tabular}
\end{center}
\end{table}

\begin{table}[htbp]
\begin{center}
\caption{The coefficient $h_{gw}$ depends on $k_g$, the
thermal conductivity of gas, $R_e$ the Reynolds number $R_e=D_D v_g/\nu_g$, on the  
rotational Reynolds number $R_{e,\omega}=D_D^2 \omega / 2\nu_g$  ($\omega$ is the angular velocity of the kiln)
and on $D_h$ an effective diameter. The latter quantity is determined from $D_D$ and $\varepsilon_B$, the angle whose sinus is equal to the ratio $2\delta/D_D$, where $\delta$ is the distance between the center of the kiln and the free surface of the granular bed (see the two last sketches of the table). The constants $A_{gw}$, $\eta_{gw}$, $B_{gw}$ and $\eta_{\omega,gw}$ are respectively equal to $0.02$, $0.93$, $8.5\times 10^{-6}$ and $1.45$. 
}\label{tab:hgw}
\begin{tabular}{|>{\centering\arraybackslash}m{.12\textwidth} | >{\centering\arraybackslash}m{.72\textwidth} |}
\hline
\multicolumn{2}{|c|}{Convective transfer between the gas and the walls}\\
\hline
\vspace*{0.8cm}Coefficient\vspace*{0.8cm}&$\displaystyle{h_{gw}=\frac{k_g}{D_h}\left(A_{gw}\,{R_{e}}^{\eta_{gw}}+B_{gw}\,{R_{e,\omega}}^{\eta_{\omega,gw}} \right)}$ with $\displaystyle{D_h={D_t}\frac{2\pi-\varepsilon_B+\sin\varepsilon_B}{2\pi-\varepsilon_B-\sin\varepsilon_B}}$ \\
\hline
Validity range&  $1100 < R_{e,\omega} < 58000$  and
$0<R_{e}<30000$ 
\cite{SeghirOuali_IntJournThermSc_2006}\\
\hline
Sketch&\vspace{0.1cm} \includegraphics*[width=0.5\linewidth]{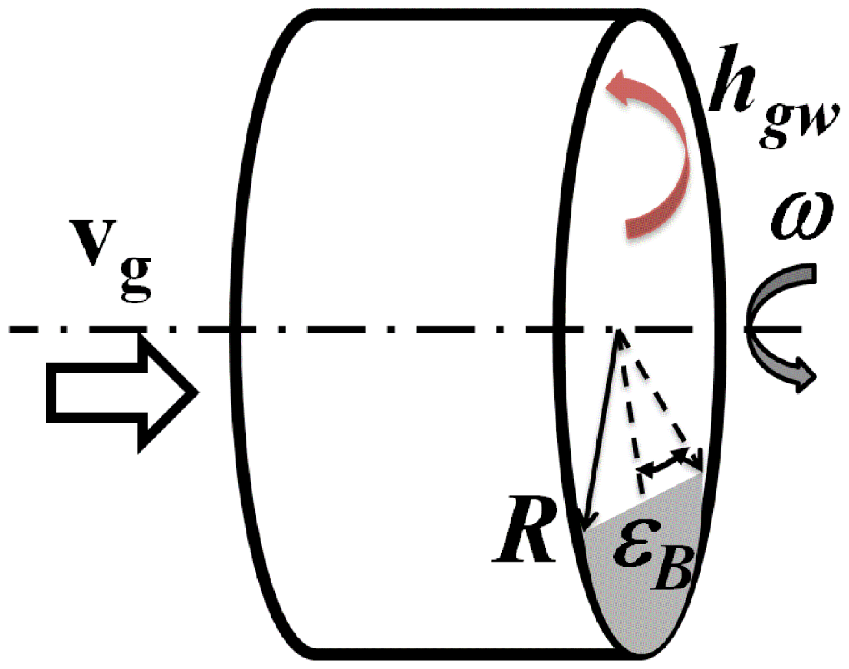}\\ 
\hline
\end{tabular}
\end{center}
\end{table}

\begin{table}[htb]
\begin{center}
\caption{The coefficient $h_{sw}$ depends on $k_s$ the
thermal conductivity of solids $D_D$ the kiln diameter, $\varepsilon_B$, $\omega$ and $a_s$, the thermal diffusivity of the solid phase, {i.e.} $k_s/(\rho_s\,C_{p,s}).$ The constants $K_{sw}$ and $\eta_{sw}$ are respectively equal to $11.6$ and $0.3$
}\label{tab:hsw}
\begin{tabular}{|>{\centering\arraybackslash}m{.12\textwidth} | >{\centering\arraybackslash}m{.68\textwidth} |}
\hline
\multicolumn{2}{|c|}{Convective transfer between the granular bed and the walls}\\
\hline
\vspace*{0.4cm}
Coefficient\vspace*{0.4cm}&$\displaystyle{h_{sw} = \frac{k_s}{\varepsilon_B D_D} K_{sw} \left( \frac{2\varepsilon_B\,D_D^2\omega}{2\,a_s} \right)^{\eta_{sw}}}$  \\
\hline
Validity range& $\omega\in[3.5, 10]\mbox{ rad.s}^{-1}$~\cite{Tscheng_CanJourChemEng_1979}\\
\hline
Sketch&\vspace{0.1cm} \includegraphics*[width=0.5\linewidth]{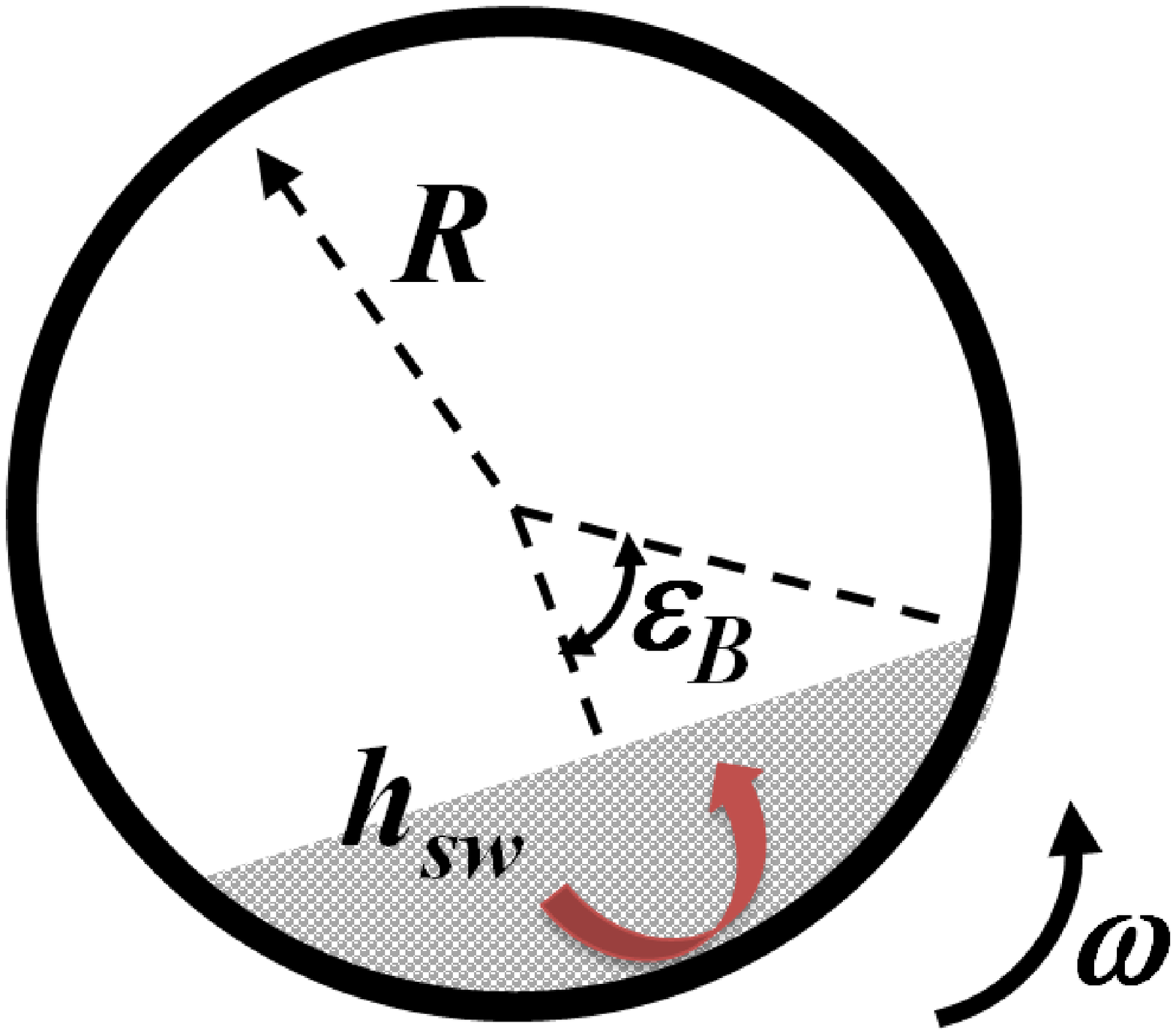}\\ 
\hline
\end{tabular}
\end{center}
\end{table}

\begin{table}[htbp]
\begin{center}
\caption{The lengths $l_{pw}$, $l_{sp}$, $l_{cg}$ and $l_{bg}$ depend on the number of baffles, $n_{F}$, the kiln diameter $D_D$ and on the angle of the bed in the kiln  
$\varepsilon_B$ (see \cite{Piton_AppliedThermalEng_2015}) which is determined using $\alpha_{B} F_t = (\varepsilon_B-\sin\varepsilon_B\cos\varepsilon_B)/\pi$ where $\alpha_B$ is the fraction of grain in the bed and $F_t$ the kiln volume fraction.}\label{tab:l}
\begin{tabular}{|c|c|}
\hline
$l_{gw}$ & $D_D \pi\varepsilon_B $\\
\hline 
$l_{bw}$ & $D_D \varepsilon_B$\\
\hline
$l_{cg}$ & $n_{F} 3\pi \left(\frac{\rho_b}{\rho_s}\right)\left(\frac{D_D^2}{2d_p}\right) F_t \alpha_{cg} $\\
\hline
$l_{bg}$ & $D_D \sin \varepsilon_B$\\
\hline
\end{tabular}
\end{center}
\end{table}

%
{Regarding the heat transfer capacities we will use $C_{p,s}=830\,\rm{  J.kg}^{-1}\rm{.K}^{-1}$ and $C_{p,g}=1100\,\rm{  J.kg}^{-1}\rm{.K}^{-1}$ which also correspond, as mentioned above,  to the asphalt plant case.}
\begin{figure}[htbp]
\begin{center}
\includegraphics*[width=\figW]{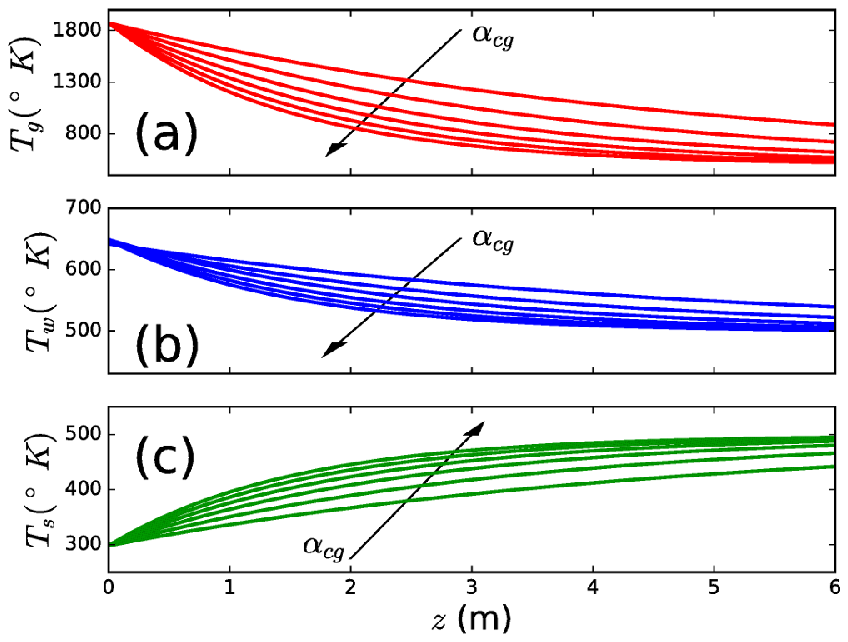}
\caption{
{Profiles} of the gas (a), wall (b) and grain (c) temperatures  versus the position $z$ along the drum axis depend on the density of grains within the granular suspension. The curves reported have been obtained for $\alpha_{cg}= 1\%$, $2\%$, $3\%$, $4\%$, $5\%$  and $6\%$. The arrows indicate increasing $\alpha_{cg}$.}\label{fig:T_z_mod}
\end{center} 
\end{figure}

Figure~\ref{fig:T_z_mod} reports variations of the temperatures versus $z$ for several values of $\alpha_{cg}$. 
{The model predicts that the temperature reached at equilibrium (for $z$ tends towards $+\infty$) is the average of the gas and grain inlet temperatures weighted by the corresponding heat flow rates}. Note that this equilibrium temperature does not depend on the exchange surfaces. The latter quantities have only an influence on the dynamics and, obviously, the distance necessary to reach the equilibrium temperature are smaller for larger exchange surfaces. This result can be obtained by a simple application of the first law of thermodynamics.\\
In order to simplify our analysis, it is reasonable to consider a kiln with a good efficiency, {i.e.} a kiln for which  the thermal transfers with the walls are much lower than those between the gas and the grains. 
This assumption is fully valid for kilns having baffles which ensure a good heat transfer between the grains and the gas.
In such a case, which corresponds to recent experimental  measurements~\cite{LeGuen_ATE_2013}, the term corresponding to the thermal exchange with the walls ($C\,D/(C+D)$) in Equation~(\ref{eqn:Lambda}) is between 2 and 30 times smaller than those corresponding to the transfer between the gas and the grains ($A$ and $B$) and the characteristic length can be reduced to the following:
\begin{equation}
\Lambda_{app}\approx\frac{E}{(I+1)(A+B)}=\frac{\kappa}{h_{cg}l_{cg}+h_{bg}l_{bg}}, \label{eqn:Lambda_approx}
\end{equation}
where $\kappa$ is the effective mass flow rate $\kappa=\dot{m_g}C_{p,g} \dot{m_s}C_{p,s} /(\dot{m_g}C_{p,g} + \dot{m_s}C_{p,s})$.
\begin{figure}[htbp]
\begin{center}
\includegraphics*[width=\figW]{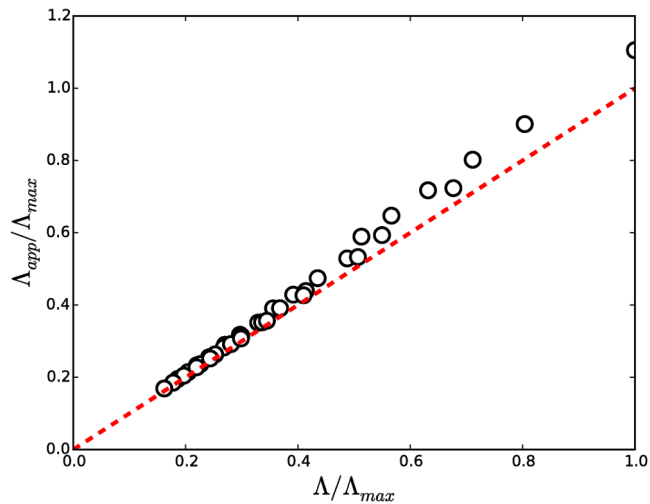}
\caption{
{The approximated characteristic length  $\Lambda_{app}$ [Equation~(\ref{eqn:Lambda_approx})] is found to be very close to the exact value $\Lambda$ [Equation~(\ref{eqn:Lambda})] demonstrating the relevancy of the latter approximation for well-insulated kilns. The points correspond to simulations obtained with $\alpha_{cg}=1\%$, $2\%$, $3\%$, $4\%$, $5\%$ and  $6\%$ and $\dot m_g= 1.1$, $1.66$, $2.06$, $2.56$, $3.12$, $3.74$ and $3.90$ kg.s$^{-1}$. The lengths are made dimensionless by $\Lambda_{max}$, the maximal value of  $\Lambda$. The dashed line corresponds to $\Lambda_{app}/\Lambda_{max}=\Lambda/\Lambda_{max}$}.}\label{fig:Lambda_approx}
\end{center} 
\end{figure}

Figure~\ref{fig:Lambda_approx} reports the evolution of the characteristic length  determined by neglecting the heat transfer with walls $\Lambda_{app}$ [Equation~(\ref{eqn:Lambda_approx})]   versus its full expression [Equation~(\ref{eqn:Lambda})]. As expected both quantities are close to each other (less that 15\% of difference) showing the relevancy of Equation(\ref{eqn:Lambda_approx}) in the case of efficient kilns. 
{Note that the approximation is especially good for large values of $\alpha_{cg}$ {i.e.} for an important  heat transfer between the granular suspension and the gas (with respect to heat transfer with the walls) and thus for small values of $\Lambda$.}\\

Obviously, and in agreement with Figure~\ref{fig:T_z_mod}, increasing $\alpha_{cg}$ and thus  the surface exchanges leads to a decrease of the characteristic length of the temperature variation along the drum. However, one should keep in mind that the present model is only valid for dilute granular suspensions, {i.e.} low values of $\alpha_{cg}$. Otherwise, the momentum exchange between the gas and the grains cannot be neglected.\\
{The characteristic length $\Lambda$ thus depends on 
{the gas and solid flow rates, $\dot{m_g}$ and $\dot{m_s}$}.
If one of the   flow rates is increased while keeping the other flow rate and all the other parameters constant, $\kappa$ and consequently $\Lambda$ increase. Thus a longer drum is required to reach the same equilibrium temperatures. 
Such a result can be understood easily. 
For example, if $\dot{m_s}$ is increased while the other parameters are kept constant, 
the amount of heat transferred from the hot gas is shared by a larger number of grains, 
thus their temperature at the outlet of the drum is lower.\\
\begin{figure}[htbp]
\begin{center}
\includegraphics*[width=\figW]{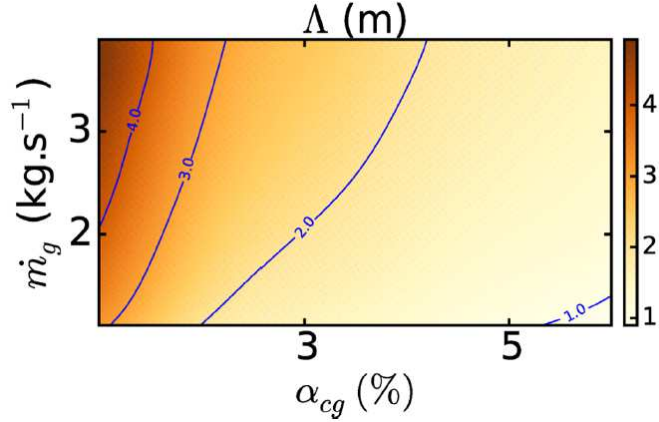}
\caption{When radiation is not taken into account, the evolution of the temperatures along the kiln are given by exponential functions whose characteristic length varies with $\alpha_{cg}$ and $\dot m_g$.}\label{fig:lambda_alpha_deb}
\end{center}
\end{figure}
The effect of both $\alpha_{cg}$ and $\dot m_g$ is highlighted on Figure~\ref{fig:lambda_alpha_deb} where we have reported the evolution of the characteristic length $\Lambda$
versus those two parameters. As expected, $\Lambda$ decreases with (i) increasing $\alpha_{cg}$ and (ii) decreasing $\dot m_g$. Thus, the smaller characteristic lengths are obtained for 
important values of $\alpha_{cg}$ and small values of $\dot m_g$.
Note however, that the  effect of $\dot{m_s}$, contrary to that of $\dot{m_g}$, might be more complex since its modification  might also induce
a variation  of $\alpha_{cg}$ and thus of the exchange lengths. To fully understand the  effect of $\dot{m_s}$ on $\Lambda$, an investigation of variety of parameters  (velocity of the drum, number of grains present in the baffles\ldots)  which depend on the details of the kiln geometry is required. Such a study is beyond the scope of this paper.}  

\section*{QUANTIFYING THE IMPORTANCE OF THERMAL RADIATION}\label{sec:quanti}
In previous analytical solution of the model,  we have neglected thermal radiation. However, it is possible to  numerically  solve the full model (including thermal radiation, see Equations~(\ref{eqn:Ts_rad}) and (\ref{eqn:Tg_rad})) following the work of Piton et al.~\cite{Piton_AppliedThermalEng_2015}.\\
\begin{figure}[htbp]
\begin{center}
\includegraphics*[width=\figW]{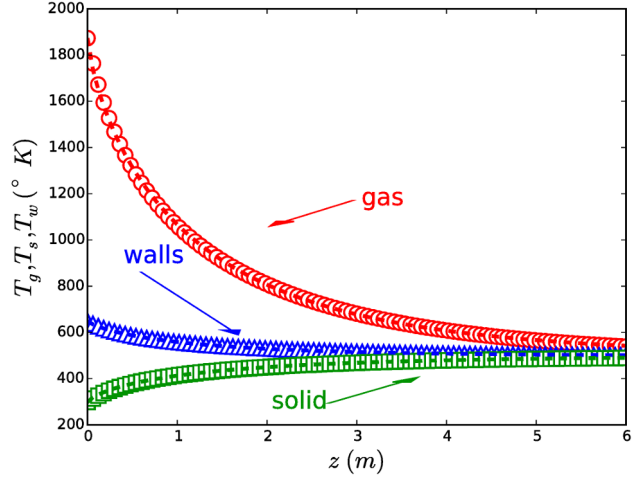}
\caption{
{Profiles} of the gas ($T_g$), wall ($T_w$) and grain ($T_s$) temperatures versus the position along the drum axis in the presence of radiation radiation (symbols). The initial gas and solid temperatures are respectively equal to $1873\,\rm K$ and $298.15\,\rm K$. The curves have been obtained using $\alpha_{cg}=3\%$. The dashed lines correspond to the fits obtained using stretched exponentials [Equation~(\ref{eqn:stretchexp}~)].}\label{fig:T_fit_stretch}
\end{center} 
\end{figure}
%
Figure~\ref{fig:T_fit_stretch} reports the numerical results obtained for $\alpha_{cg}=3\%$. 
Note that, as expected, the heat supplied to the system at its outlet is the same with and without radiation as long as the length of the drum is large enough to reach the equilibrium temperature. Indeed, since the system is insulated the temperatures at the equilibrium, {i.e.} $T_g(\infty)=T_{g,\infty}$, $T_s(\infty)=T_{s,\infty}$ and $T_w(\infty)=T_{w,\infty}$, are also the same, independently  of the fact that radiation is taken --or not-- into account.\\
As mentioned above,  the general equations which include thermal radiation terms cannot be solved analytically due to the $T^4$-terms. 
In such a case, to define a characteristic length of the phenomena and thus obtain an efficiency criterion 
it is necessary to fit the numerical results by an adequate function.
{  The  $z$-profiles of the temperatures obtained numerically cannot be fitted by an exponential function whose variation is too rapid. Thus, following what is classically done in such a case, we try to fit our results by a stretched exponential given by Equation~(\ref{eqn:stretchexp}) and obtained satisfactory results.
}   
First used by Kohlrausch in 1854~\cite{Kohlrausch1854} to describe mechanical creep, this
expression was far later popularized by Williams and Watts~\cite{Williams1970} who described dielectric relaxation in polymer as being stretched exponential functions. It is now frequently applied
to a large range of relaxations in disordered thermal systems as glasses or granular materials \cite{Richard_NMAT_2005,Ribiere2005d} and is often called the KWW law. Its expression is given by: 
\begin{equation}
T_\gamma(z) = T_{\gamma,\infty} + \left(T_{\gamma,0} - T_{\gamma,\infty}\right)\exp\left(-(z/\lambda_\gamma)^{\beta_\gamma}\right),
\label{eqn:stretchexp}
\end{equation}
where index $\gamma$ stands for the studied phase ({i.e.} solid, gas or wall).
Note that  the stretched exponential is also known as the complementary cumulative Weibull distribution. 
{To explain the use of such a fit function, one can consider a small region around  of the drum $z=z_i$  for which the radiative terms $E_{ij}l_{ij}(T_g^4-T_s^4)$ can be approximated by
$E_{ij}l_{ij}(T_g-T_s)(T_{g,i}+T_{s,i})(T_{g,i}^2+T_{s,i}^2)$ where $T_{g,i}$ and $T_{s,i}$ are respectively the solid and gas temperature at $z=z_i$. This assumption leads to a linear system of differential equations around $z_i$. The solution drives  exponential functions whose characteristic length depends on $T_{g,i}$ and $T_{s,i}$ and increases with $z_i$. The full solution may thus be seen as a combination of simple exponential decays which is known as a stretched exponential function.}\\

\begin{figure}[htbp]
\begin{center}
\includegraphics*[width=\figW]{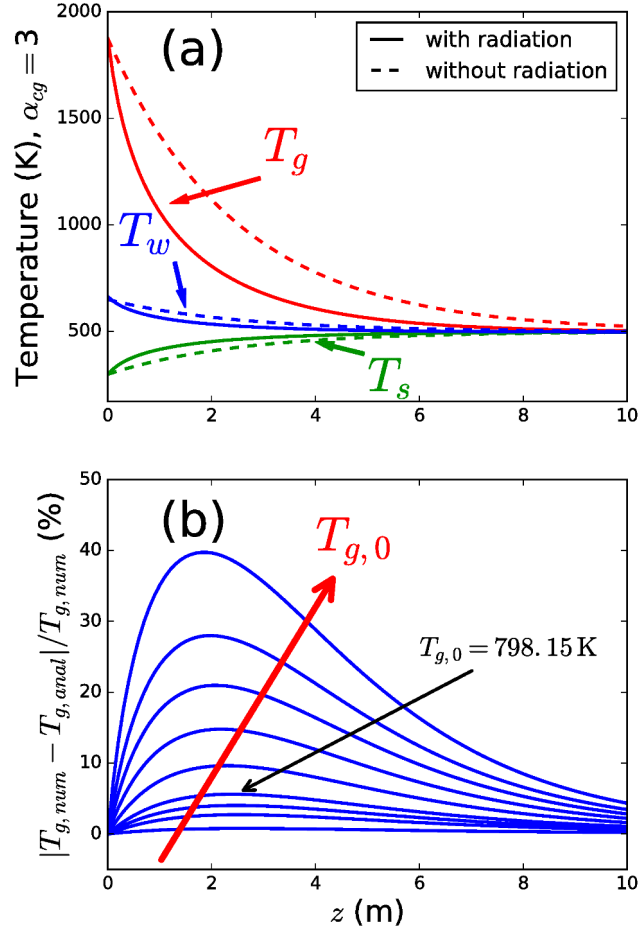}
\caption{
(a) 
{Profiles} of the gas ($T_g$), wall ($T_w$) and grain ($T_s$) temperatures versus the position along the drum axis with (numerical) and without (analytical) radiation. The initial gas and solid temperatures are respectively equal to $1873\, K$ and $298.15\, K$. The curves have been obtained using $\alpha_{cg}=3\%$. As expected the 
{variations} are more important with radiation and, at thermal equilibrium, all the temperatures are identical.
(b) Relative difference between the numerical and analytical temperatures along the drum axis at different initial gas conditions for $\alpha_{cg}=3\%$. At the inlet, the grain temperature is fixed to $298.15\, \rm K$. The curves have been obtained for 
$398.15\, \rm K$, $598.15\, \rm K$, $698.15\, \rm K$, $798.15\, \rm K$, $998.15\, \rm K$,
 $1198.15\, \rm K$, $1398.15\, \rm K$, $1598.15\, \rm K$, and $1898.15\, \rm K$. The red arrow indicates increasing $T_{g,0}$.}\label{fig:effet_T4_bis}
\end{center} 
\end{figure}

As mentioned above the temperatures reached for $z\rightarrow +\infty$ are the same with and without radiation : $T_{g,\infty}=T_{s,\infty}=(T_{g,0}I+T_{s,0})/(I+1).$
The expressions of the gas and solid temperatures can thus be respectively written as :
\begin{equation}\left\{
\begin{array}{ccc}
T_g &=& \frac{T_{g,0}-T_{s,0} }{I+1 }\exp\left(- \left(z/\lambda_g\right)^{\beta_g}\right)+\frac{T_{g,0}I+T_{s,0}}{I+1},\\
\ & & \\
T_s&=&-I \frac{T_{g,0}-T_{s,0} }{I+1 }\exp\left(- \left(z/\lambda_s\right)^{\beta_s}\right)+\frac{T_{g,0}I+T_{s,0}}{I+1}.
\end{array}
\right.\label{eqn:sol_exp_etiree}\end{equation}
{
It should be pointed out that, for the solid phase, the stretched exponential fit  is not able to reproduce correctly the data close the equilibrium value of the temperature. 
This is potentially problematic for very long drum and/or very low values of $\alpha_{cg}$ ({i.e.} $\alpha_{cg} \leq 0.1\%$). However, close to the inlet of the kiln, the fit is fully valid.} 
The two free parameters used for the fit are  $\lambda_\gamma$, the characteristic length of the stretched exponential and $\beta_\gamma$, its power. 
{Interestingly, our numerical results show that the dependence of the phase is very weak, {i.e.}  $\beta_s\approx \beta_g \approx \beta_w = \beta$ and $\lambda_s\approx \lambda_g \approx \lambda_w = \lambda$ are reasonable approximations as long as we restrict ourselves to the region of the kiln for which the stretched exponential fit is valid.}
\begin{figure}[htbp]
\begin{center}
\includegraphics*[width=\figW]{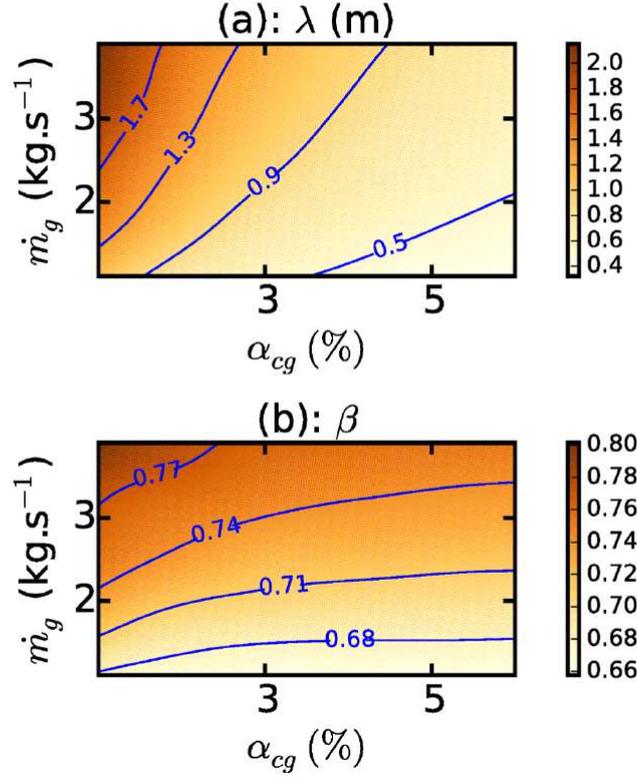}
\caption{When radiation is taken into account, the temperatures are well fitted by stretched exponential functions whose parameters (the characteristic length $\lambda$ (a) and the power $\beta$ (b)) are both found to be independent of the phase but vary with $\dot m_g$ and $\alpha_{cg}$.}\label{fig:beta_lambda_alpha_deb}
\end{center}
\end{figure}
Therefore the 
{profiles}  of the temperatures within the system are determined by the two parameters of the stretched exponential which depend on $\alpha_{cg}$ and $\dot m_g$ (see Figure~\ref{fig:beta_lambda_alpha_deb}).\\
Two important remarks should be made. 
Firstly, both $\lambda$ and $\beta$ increase with the gas flow rate $\dot m_g$. 
{This result indicates that increasing the mass flow rate of the gas requires to use longer kilns to reach equilibrium. 
Secondly, an increase of $\alpha_{cg}$ leads to a decrease of the exponent of the stretched exponential and of the characteristic length.
The latter result is expected since an increase of $\alpha_{cg}$ increases the surface exchange between the granular suspension and the gas and thus improves the heat exchange between the gas and the grains. 
Note that the relation between the two parameters $\beta$ and $\lambda$ is not one-to-one. In other words one cannot deduce one of the parameters from the knowledge of the other.
However, an interesting result should be pointed out: the relation between the two parameters $\lambda$ and $\beta$ are one-to-one for each value of
$\alpha_{cg}$, {i.e.}, the two parameters $\beta$ and $\lambda$ are correlated if $\alpha_{cg}$ is kept constant. Interestingly, the correlation is linear (see Figure~\ref{fig:L_BIGL}c) 
{ but, to date,  we have no explanation for this linearity which warrants further studies.} 
Our analysis raises another question: is the characteristic length in presence of radiation, ({i.e.} $\lambda$) linked to that obtained in absence of radiation, ({i.e.} $\Lambda$)? Figure~\ref{fig:L_BIGL}b reports the evolution of the former quantity versus the latter.
Clearly and surprisingly, they are linearly correlated for each value of $\alpha_{cg}$. Naturally, $\beta$ and $\lambda$ being correlated for each value of $\alpha_{cg}$, $\beta$ is also
linked to $\Lambda$, the characteristic length in absence of radiation (Figure~\ref{fig:L_BIGL}c).
The origin of those linear correlations is not established.\\ 
\begin{figure}[htbp]
\begin{center}
\includegraphics*[width=\figW]{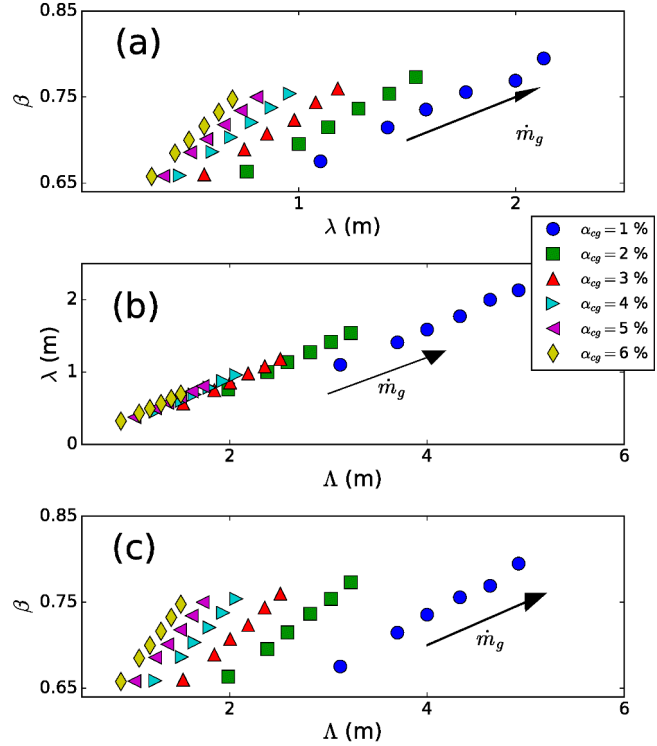}
\caption{In presence of radiation, the characteristic length obtained by fitting the temperature evolution, $\lambda$ and its power $\beta$ are linked linearly for each value of $\alpha_{cg}$, the fraction of grains in the granular suspension (a). Similarly, for each value $\alpha_{cg}$, the same characteristic length is also linked linearly to $\Lambda$ the characteristic length in absence of radiation (b).  Consequently the power $\beta$ is also linearly linked to $\lambda$, for each value of $\alpha_{cg}$. The arrows indicate increasing $\dot{m_g}$.}\label{fig:L_BIGL}
\end{center}
\end{figure}
%
{In our system, the gas temperatures range from a few hundreds of Kelvins to a few thousands, bringing us to the question of whether radiation should be neglected ({i.e.} for low temperatures) or not ({i.e.} for high temperatures).}  
{Thus, }we compare the temperatures obtained with the analytical solution of the model without radiation 
to those obtained by the numerical resolution of the full model. Figure~\ref{fig:effet_T4_bis}a reports the numerical ({i.e.} with radiation) and analytical ({i.e.} without radiation) temperature profiles for a fraction of grains belonging to the granular suspension $\alpha_{cg}=3\%$. 
As expected, the influence of 
the radiation is significant especially at the early stages of the heating, {i.e.} when the gas temperature $T_g$, and thus $T_g^4$, are important. Similarly, the influence of the radiation becomes smaller when the gas temperature $T_g$ decreases. 
Radiation might therefore be neglected only at lower temperature values. To illustrate this purpose, we show in Figure~\ref{fig:effet_T4_bis}b   
the analytical and numerical relative variations of $T_g$ ({i.e.} the quantity 
$(T_g - T_{g,\infty})/T_{g,0}$) 
for $\alpha_{cg}=3\%$ and for several  initial temperatures (see caption). As an example, for $\alpha_{cg}=3\%$ and $T_{g,0}=798\mbox{ K}$ the maximum relative error is found to be around $4\%$. }

\begin{figure}[htbp]
\begin{center}
\includegraphics*[width=\figW]{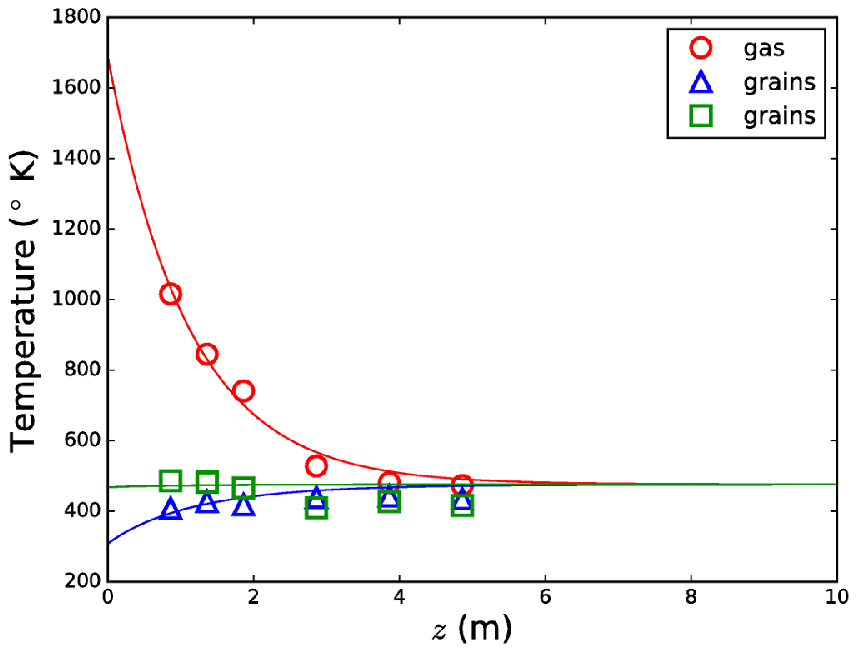}
\caption{
The temperatures obtained experimentally in an industrial rotary kiln devoted to asphalt production~\cite{LeGuen_ExTFS_2014} (symbols) are well fitted by  simple exponentials sharing the same characteristic length and equilibrium temperature. The rate of the production is $140$ Tons per hour.}\label{fig:fit_exp}
\end{center} 
\end{figure}

The validity of our model is then tested by fitting experimental measurements of the gas temperature data obtained experimentally in an industrial flight rotary kiln devoted to the asphalt materials manufacturing~\cite{LeGuen_ExTFS_2014} by simple exponential functions (Figure~\ref{fig:fit_exp}). 
The characteristic length obtained is close to that obtained by Equation~(\ref{eqn:Lambda}) (roughly $10\%$ difference).
The grain and wall temperatures are also well reproduced by simple exponentials for which the characteristic lengths and the equilibrium temperatures are set to the values obtained with the fit of the gas temperature. 
It should be pointed out that  stretched exponentials also fit  correctly experimental data, without being significantly better. Unfortunately, the number of experimental measurements is too small to differentiate the two fits.
\section*{DISCUSSION: OPTIMAL LENGTH OF THE KILN}\label{sec:eff}
The above study predicts the 
{spatial evolution} of the grain temperature within a model kiln with or without radiation. 
It is thus possible to estimate the optimal length of a  rotating kiln allowing the grains to reach a given temperature, $T_\nu$. In the present case, calculation of the requested temperature, $T_\nu$, is obtained for a given length $L_\nu$, in such way that
${T_\nu}={T_{s}(z=L_\nu)}$.

Of course, for a given grain flow rate, the gas flow rate or the initial temperatures have to be chosen in order that the equilibrium temperature
$T_{s,\infty}=\left(T_{g,0}I+T_{s,0}\right)/\left(I+1\right)$ is larger or equal to $T_\nu$. Otherwise, $T_\nu$ is unreachable.
%
If $T_\nu$ can be reached, the simplest solution consists in using a very long kiln to be sure that the outlet temperature of the grains is equal to or larger than $T_\nu$. However,  a too long drum is not efficient since energy is used to increase the grain temperature up to an unnecessarily high value. 
A criterion is therefore necessary to quantify the efficiency of the kiln relative  to its length.
From the above analytical 
{resolution} ({i.e.} without radiation) our empirical expressions ({i.e.} with radiation) of $T_g$, the length $L_\nu$ at which the grain temperature reaches $T_\nu$ can be easily obtained. 
If the length of the kiln $L_D$ is lower than $L_\nu$, the requested temperature cannot be reached. The efficiency can be therefore defined as the following ratio:
\begin{equation}
\varepsilon=\frac{\int_0^{L_D} T_s(z)\,dz}{\int_0^{L_\nu} T_s(z)\,dz}\mbox{ if }L_D<L_\nu.\label{eqn:eff_1}
\end{equation}
In such a case, the efficiency increases from zero when the length of the drum is increased and reaches its maximal value ($\varepsilon=1$) when the length of the drum is equal to $L_\nu$.
The criterion can be seen as the ratio of the energy used to set the grain temperature to its value at the outlet of the drum over the energy necessary to achieve the requested temperature.\\
\begin{figure}[htb]
\begin{center}
\includegraphics*[width=\figW]{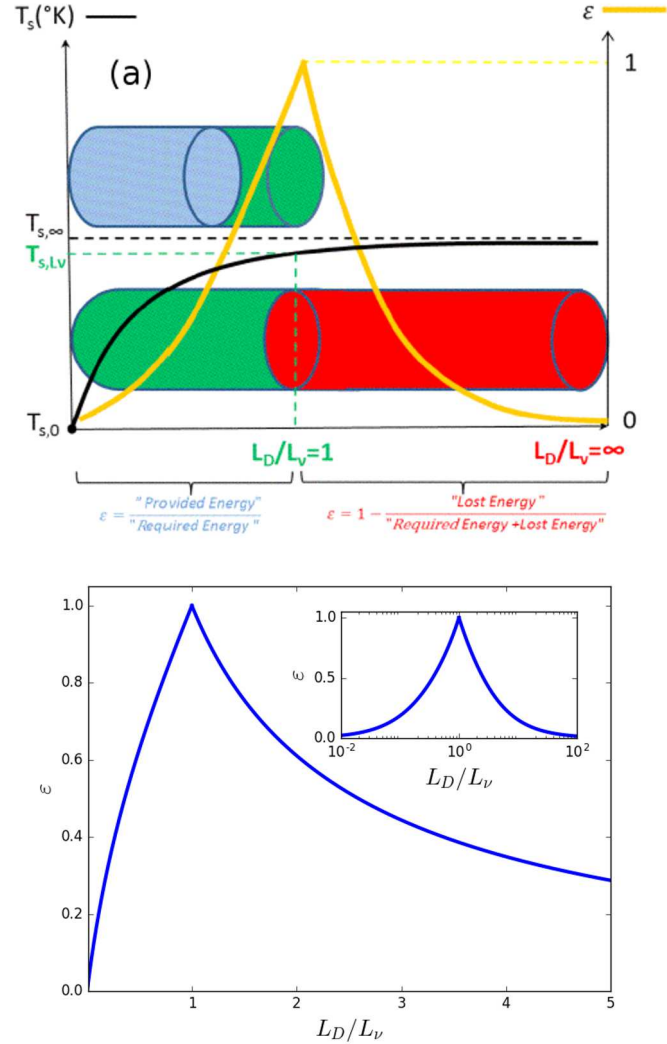}
\caption{(color online) 
(a) Efficiency criteria. The requested temperature for the grains is $T_\nu$ and the length at which that temperature is reached is $L_\nu$. This temperature must be lower than the equilibrium temperature reached for an infinite drum otherwise it is unreachable for any length.
If the size of the drum $L_D$ is lower to $L_\nu$ (blue or light grey drum), the temperature $T_{s,\nu}$ cannot be reached. The efficiency $\varepsilon$ increases towards $1$ when $L_D$ tends towards $L_\nu$ which corresponds to the ideal length of the drum (green or medium grey drum).
On the contrary if $L_D>L_\nu$ (red or dark grey drum) the grain temperature at the outlet is greater than $T_{s,\nu}$, leading to a waste of energy.
(b) The efficiency of the kiln depends on the drum's length. The present curve has been obtained using Equation~(\ref{eqn:stretchexp}) with $\beta=0.76$ and $\lambda/L_\nu=0.47$. Efficiency rises to reach its maximum value at $ L_{D}=L_{v}$. Above $L_{v}$, the energy waste increases and the efficiency decreases progressively. The inset reports the same data in a lin-log scale.
}\label{fig:eff}
\end{center}
\end{figure}
For a drum whose length is larger than $L_\nu$, the temperature of the grains at the outlet of the device is larger that the requested temperature. Thus the final temperature of the grains is higher than the requested temperature and part of the used energy is wasted. Therefore, the efficiency of the process decreases with drum's length. It can be defined as:
\begin{equation}
\varepsilon=1- \frac{\int_{L_\nu}^{L_D} T_s(z)\,dz}{\int_0^{L_D} T_s(z)\,dz}\mbox{ if }L_D>L_\nu.\label{eqn:eff_2}
\end{equation}
In the latter equation, the ratio of the two integrals can be seen as the ratio of the wasted energy over the total energy.
The definition of our efficiency criterion is sketched in Figure~\ref{fig:eff}a and an example of the dependence  on the efficiency with the length of the drum is given on
Figure~\ref{fig:eff}b. Note that radiation, by increasing heat exchange between the gas and the granular suspension, improves the efficiency of the kiln for a given $z$-position. However, the shape of the curve is not significantly influenced by the presence (or the absence of) radiation as long the length of the kiln is made dimensionless by $L_\nu$.”


{Such kind of approach can also be applied to existent kilns, {i.e.} kilns for which $L_D$ is fixed. 
In this case, the gas temperature at the inlet can be chosen to obtain the requested solid temperature at the outlet (Equations~(\ref{eqn:sol_exp}) or (\ref{eqn:sol_exp_etiree})).  Yet, if for practical reasons, it is not possible to set the inlet temperature at its optimal value, the efficiency can be determined using Equations~(\ref{eqn:eff_1}) and (\ref{eqn:eff_2}) in which $L_\nu$ is the length corresponding to the requested temperature.}

\section*{CONCLUSION}\label{sec:conclu}
In this paper, we report a theoretical approach aiming to model heat transfer between a convective gas and a model  granular system.
The studied system is typical of a rotary kiln 
commonly used in industry to dry grains (asphalt production, fertilizers production, waste treatment\ldots) {i.e.} a dilute suspension of heavy grains interacting with a convected gas only through heat exchange.
We divided the system into different ``phases'' (gas, granular suspension, granular bed, walls) which interact  through exchange surfaces.
The granular phases are treated as effective continuous  media. The model is  solved analytically in the absence of radiation and it is shown that the temperature 
{profiles along the drum are governed} by one characteristic length whose expression depends on the heat transfer properties and on the exchange surfaces. 
In the presence of radiation, the 
{profiles} 
of the temperatures  along the axis of the drum are well fitted by stretched exponentials whose parameters depend on the surface exchanges but are the same for each phase of the system. As expected, the presence of radiation does not modify the equilibrium temperature but reduces the length required to reach equilibrium.
Finally a criterion quantifying the efficiency of the process is proposed. It is based on an estimation of the optimal length of the kiln as a function of the temperature of the grains at the outlet of the kiln. \\
Similar analysis should be applied to the counter current case 
in order to obtain rigorous energetic diagnostics of different rotary kiln systems.  
It should be pointed out that several assumptions have been made to derive the present model. Although they may appear to be somewhat crude, they are justified in the case of a convected gas moving through dilute suspension of heavy grains, which corresponds to the majority of application in the field of civil engineering. However, the large variety of granular materials present in other industrial fields deserves consideration of a more detailed model which should take into account a diffusional time within the grain for larger Biot number and agitation of grains within the suspension ({i.e.} granular temperature).
{Note also that, although in the case of kiln devoted to asphalt production the moisture of the grains evaporates quickly and in vicinity of the outlet, it should probably be considered for other types of kilns (\textit{e.g.} kilns devoted to fertilizer production).}
{Our simple model can be used in industry to help the design of efficient kilns. Indeed it gives an efficiency criterion which estimates the optimal kiln length, {i.e.} the length of the kiln which allows to reach a given final temperature. Thus it helps the design kilns which are really adapted to the intended application.} 

\section*{Acknowledgments}
We thank Bogdan Cazacliu, Yannick Descantes, Nicolas Roquet, Riccardo Artoni and Erwan Hamard for fruitful discussions and Jean-Marc Paul for technical assistance. We are indebted to Andrew Hobbs for a critical reading of the manuscript and for kindly providing us with unpublished numerical results.   


\end{document}